\documentclass[preprintnumbers,twocolumn,amsmath,amssymb]{revtex4}
\usepackage{dcolumn}
\usepackage{bm}
\usepackage[dvips]{color}
\usepackage[pdftex]{graphicx}
\newcommand{\be}{\begin{equation}}
\newcommand{\ee}{\end{equation}}
\newcommand{\ba}{\begin{array}}
\newcommand{\ea}{\end{array}}
\newcommand{\bea}{\begin{eqnarray}}
\newcommand{\eea}{\end{eqnarray}}
\evensidemargin=\oddsidemargin

\begin{document}

\title{\large 
$B_{d,s}^0-\bar{B}_{d,s}^0$ mixing and Lepton Flavour Violation in SUSY GUTs:
{impact of the first measurements of $\phi_s$}}

\author{J.K.Parry$^{1}$\footnote{jkparry@tsinghua.edu.cn}, 
Hong-Hao Zhang$^{2}\footnote{zhh98@mail.sysu.edu.cn}$
}

 \affiliation{ $^{1}$
       Center for High Energy Physics,
       Tsinghua University, Beijing 100084, China \\
$^{2}$ School of Physics \& Engineering,
Sun Yat-Sen University, Guangzhou 510275 China
}

 \begin{abstract}
In this work we re-examine the correlation 
between $B_{d,s}^0-\bar{B}_{d,s}^0$ mixing
and Lepton Flavour Violation in the light of recent experimental measurements
in the $B_s$ system. 
We perform a generic SUSY analysis of the allowed down 
squark mass insertion parameter space. 
In the SUSY GUT scenario this parameter
space is then used to make predictions for LFV branching ratios. 
We find that the recent measurement 
for the CP phase $\phi_s$ excludes the lowest rates for $\tau\to\mu\gamma$
and provides a lower bound of $\sim 3\times 10^{-9}$ for $\tan\beta=10$.
Future experimental improvements in the bound on $\tau\to\mu\gamma$ 
and the measurement of $\phi_s$ 
will constitute a strong test of the SUSY GUT scenario.
 \\[2mm]
PACS:  12.10.-g; 12.60.Jv; 13.35.Dx; 14.40.Nd; 13.20.He \hfill
Nucl.Phys.B802:63-76,2008
 \end{abstract}

 \maketitle

\section{Introduction}

In recent years there have been great efforts made 
in the search for New Physics (NP) in the B system which 
has been seen as an extremely promising place to find
NP effects as deviations of flavour-changing
neutral-currents (FCNC) from their Standard Model(SM)
expectations. 
The first measurements of $B_s\!-\!\bar{B}_s$ mixing have been
observed at $\rm D\emptyset$ \cite{hepex0603029} and CDF
\cite{hepex0609040} experiments reporting a mass difference,
\bea
&&17\,{\rm ps}^{-1}<\Delta M_s < 21\,{\rm ps}^{-1}\hspace{1cm}
(90\%\,{\rm C.L.} \, \,\mathrm{{\rm D}\emptyset})\\
&& \Delta M_s=17.77\pm 0.10\pm 0.07 \, {\rm ps}^{-1}\hspace{1cm}(\mathrm{CDF})
\label{dms}
\eea
The constraints from eq.~(\ref{dms})
strongly restrict the allowed parameter space for the NP contribution
to $B_s\!-\!\bar{B}_s$ mixing 
\cite{hepph0604249,Parry:2006vq,Lenz:2006hd,Freitas:2007dp,hepph0703214,Trine:2007ma,Dutta:2006gq,Parry:2005fp,Parry:2006mv}.

The first measurement of the CP phase
associated with $B_s\!-\!\bar{B}_s$ mixing, $\phi_s$, was
reported by the ${\rm D}\emptyset$ 
collaboration \cite{hepex0702030}.
More recently, flavour-tagged measurements have
now been made by both
the CDF collaboration \cite{Aaltonen:2007he} and 
the ${\rm D}\emptyset$ collaboration \cite{:2008fj}. 
Combining these latest results 
the UTfit group \cite{Bona:2008jn} found that
the new physics phase deviates from zero by about $3\sigma$,
\bea
\phi_s=(\phi_{s}^{NP}-2\beta_s)
\eea
with $\beta_s=0.018\pm0.001$ and,
\bea
\phi_{s}^{NP}=\left[-60.9,\,-18.58\right]^{\rm o}\cup
\left[-156.90,\,-106.40\right]^{\rm o}
\label{phis_av}
\eea
at $95\%$ C.L. There still remains a two fold ambiguity
in the measurement due to the symmetry in 
$\phi_s\to (\pi - \phi_s)$.

In the $B_d$ system the measured mass difference given by the
``Heavy Flavour Averaging Group''(HFAG)\cite{hepph0603003} is,
\bea
\Delta M_d=0.507\pm 0.004\, {\rm ps}^{-1}
\label{dmd}
\eea
and the phase, $\phi_d$, has been measured as,
\bea
\phi_d=0.757\pm 0.044
\label{phid}
\eea

A very appealing model of NP is the supersymmetric grand 
unified theory (SUSY GUT).
In such models there is a deep
underlying connection between quarks and leptons above the 
GUT scale. For example in SU(5) the singlet down quarks 
are related to the lepton doublet as they live in the 
same $\bf{\bar{5}}$ representation. This deep relation 
between down quarks and leptons is broken at the GUT scale
but its presence may still be felt all the way down at the 
Electroweak scale. As a result of this relationship the 
right-handed down squark and left-handed slepton 
mixings are related to each other. 
In such SUSY GUTs there is 
hence a deep connection between LFV rates such as BR$(\tau\to\mu\gamma)$
and FCNCs such as $B_s\!-\!\bar{B}_s$ mixings and its associated
CP phase \cite{hepph0702144,hepph0702050,hepph0205111,hepph0007328}.

In this work we re-examine the correlation between FCNCs and 
LFV rates in SUSY GUTs. Specifically we are interested in the 
connection between the allowed NP parameter space of 
$B_s\!-\!\bar{B}_s$ mixing and the rate of $\tau\to\mu\gamma$.
To this end we first perform a generic SUSY analysis,
of the allowed parameter space of down squark mass insertions
$\delta^d_{RR}$. Our analysis in generic in that no 
particular SUSY-breaking mechanism shall be assumed.
Having determined the allowed parameter space 
we then exploit the GUT relationship of squark and slepton 
mass insertions to make realistic predictions for 
the rates of $\tau\to\mu\gamma$, $\tau\to e \gamma$ and 
their ratio. 
Recently a similar study was undertaken \cite{hepph0702050}
on the correlation between LFV and $B_q$ mixing. 
We believe that the NP contribution to $B_d$ mixing was 
under-estimated in that work leading to over-estimates
for the branching ratio of $\tau\to e\gamma$. 
In this work we use an analytic form for the gluino contribution
to $B_d$ mixing derived in appendix A in the mass insertion
approximation. This allows us to 
correctly compute the NP contribution to both $B_d$ and $B_s$ mixing.
Furthermore the recent measurement of the CP phase $\phi_s$ has prompted a 
re-examination of this correlation. Crucially we find that these
recent measurements severely restrict the allowed parameter
space and provide a lower bound for the rate of $\tau\to\mu\gamma$.
Finally, we explore an example of an SO(10) inspired model
and discuss the tension between the constraints of LFV and FCNCs.

\section{Constraints on the SUSY contribution to $B_q$ mixing}

The $\Delta B=2$ transition between $B_q$ and $\bar{B}_q$ mesons is 
defined as,
\bea
\langle B^0_q | {\mathcal{H}}^{\Delta B=2}_{eff} |\bar{B}_q^0  \rangle
=2M_{B_q}M^q_{12}
\eea
where $M_{B_q}$ is the mass of the $B_q$ meson. We can then define the 
$B_q$ mass eigenstate difference as,
\bea
\Delta M_q\equiv M_H^q-M_L^q=2|M_{12}^q|
\eea
and its associated CP phase,
\bea
\phi_q=arg(M_{12}^q)
\eea
In the Standard Model $M_{12}^q$ is given by,
\be
M_{12}^{q,{\rm SM}}=\frac{G_F^2 M_W^2}{12 \pi^2}
M_{B_q}\hat{\eta}^B\,f_{B_q}^2 \hat{B}_{B_q}
(V_{tq}^* V_{tb})^2\, S_0(x_t)
\ee
where $G_F$ is Fermi's constant, $M_W$ the mass of the W boson,
$\hat{\eta}^B=0.551$ is a short-distance QCD correction identical 
for both the $B_s$ and $B_d$ systems. The bag parameter $\hat{B}_{B_q}$ 
and decay constant $f_{B_q}$ are non-perturbative quantities and contain
the majority of the theoretical uncertainty. $V_{tq}$ and $V_{tb}$ are elements
of the Cabibbo-Kobayashi-Maskawa (CKM) matrix, 
and $S_0(x_t\equiv \bar{m}_t^2/M_W^2)=2.32\pm 0.04$.

The NP contribution to 
$B_q$ mixing may be parameterized
in a model independent way as,
\bea
\Delta M_q&=&\Delta M_q^{\rm SM}\,|1+R_q| \label{totalM}\\
\phi_q&=&\phi_q^{\rm SM}+\phi_q^{\rm NP}=
\phi_q^{\rm SM}+arg(1+r_q\, e^{i\sigma_q})
\label{totalphi}
\eea
where $R_q\equiv r_q\, e^{i\sigma_q}=M_{12}^{q,{\rm NP}}/M_{12}^{q,{\rm SM}}$ 
denotes the relative size of the 
NP contribution. The dominant 
SUSY contribution to $B_q$ mixing comes from the gluino 
contribution which may be written as,
\bea
R^{\tilde{g}}_q&=&
a^q_1(m_{\tilde{g}},x)\,
\left[(\delta_{q3}^d)^2_{RR}+(\delta_{q3}^d)^2_{LL}\right]
\nonumber\\
&&+a^q_4(m_{\tilde{g}},x)\,
(\delta_{q3}^d)_{LL}(\delta_{q3}^d)_{RR}+\ldots
\label{Rg}
\eea
Here we have ignored terms proportional to $\delta^d_{RL,LR}$
mass insertions as they are expected to be heavily suppressed
due to constraints from $b\to s\gamma$ \cite{hepph0702144}. 
In appendix A we give details of the functions $a^q_1$ and $a^q_4$.

We can now constrain both the magnitude and phase of the NP
contribution, $r_q$ and $\sigma_q$, through the comparison of 
the experimental measurements with SM expectations. From the 
definition of eq.~(\ref{totalM}) we have the constraint,
\bea
\rho_q\equiv \frac{\Delta M_q}{\Delta M_q^{\rm SM}}
=\sqrt{1+2r_q\cos\sigma_q + r_q^2}
\label{rho}
\eea
for $r_q$ and $\sigma_q$. 
The values for $\rho_q$ given by the UTfit analysis 
\cite{Bona:2008jn,Bona:2007vi} at the $95\%$ C.L. are,
\bea
\rho_d&=&\left[0.53,\,2.05\right]\\
\rho_s&=&\left[0.62,\,1.93\right]
\eea

From eq.~(\ref{totalphi})
the phase associated with NP can also be 
written in terms of $r_q$ and $\sigma_q$,
\bea
\sin\phi_q^{\rm NP}&=&\frac{r_q\sin\sigma_q}
{\sqrt{1+2 r_q\cos\sigma_q+r_q^2}},
\nonumber\\
\cos\phi_q^{\rm NP}&=&\frac{1+r_q\cos\sigma_q}
{\sqrt{1+2 r_q\cos\sigma_q+r_q^2}}
\label{phiNP}
\eea
Here \cite{Bona:2008jn,Bona:2007vi} gives the $95\%$ C.L. constraints,
\bea
\phi_d^{\rm NP}&\hspace{-1mm}=&\hspace{-1mm}
\left[-16.6,\,3.2\right]^{\rm o}\\
\phi_{s}^{NP}&\hspace{-1mm}=&\hspace{-1mm}
\left[-60.9,\,-18.58\right]^{\rm o}\cup
\left[-156.90,\,-106.40\right]^{\rm o}
\label{phisdNP}
\eea

In order to consistently apply the above constraints all input 
parameters are chosen to match those used in the analysis
of the UTfit group \cite{Bona:2008jn,Bona:2007vi} with the 
non-perturbative parameters,
\bea
f_{B_s}\hat{B}_{B_s}^{\frac{1}{2}}&=&262\pm 35 \,{\rm MeV}\\
\xi&=&1.23\pm 0.06 \,{\rm MeV}\\
f_{B_s}&=&230\pm 30 \, {\rm MeV}\\
f_{B_d}&=&189\pm 27 \, {\rm MeV}
\eea

Now we have a clear picture of the constraints to be imposed 
on the $r_q-\sigma_q$ plane as listed in eq.~(\ref{rho}-\ref{phisdNP}).
In the following sections these constraints shall be used 
in a generic SUSY analysis of the mass insertion parameter space,
followed by a study of the correlation of $B_q$ mixing and LFV 
rates in the minimal SUSY GUT.

\section{Correlation between $B_q$ mixing and LFV rates in SUSY GUTs}

In many ways the 
prototypical GUT is SU(5) as it is the minimal group which can unify
the gauge group of the Standard Model. In the SU(5) GUT the quarks
and leptons are placed into ${\bf 10}=(Q,\, u^c,\, e^c)$ and 
${\bf \bar{5}}=(L,\, d^c)$ representations. 
Due to the symmetry of this simple SUSY GUT
there exists relations amongst the slepton and squark soft 
SUSY breaking masses,
\bea
m_{10}^2=m_{\tilde{Q}}^2=m_{\tilde{U}}^2=m_{\tilde{E}}^2,
\hspace{0.5cm}
m_{5}^2&=m_{\tilde{L}}^2=m_{\tilde{D}}^2
\label{GUT}
\eea
These relations hold for scales at and above the GUT scale. 
Interestingly this implies that left-handed slepton mixing and 
right-handed down squark mixing are related. 
This relation can still be felt
very strongly at the Electroweak scale in the correlation of 
LFV rates and FCNCs.


Further renormalization group(RG) 
evolution down to the Electroweak scale has very 
little effect on the size of these off-diagonal elements \cite{hepph0702144}.
So we may  assume that these GUT scale values are approximately 
equal to their values at the electroweak scale. 
Hence it is a fair approximation
to assume that $(m_{\tilde{D}}^2)_{ij}\simeq (m_{\tilde{L}}^2)_{ij}$
at the electroweak scale. Then the 
$(\delta^d_{ij})_{RR}\equiv (m_{\tilde{D}}^2)_{ij}/m_{\tilde{q}}^2$
contributions
to FCNC and 
$(\delta^l_{ij})_{LL}\equiv (m_{\tilde{L}}^2)_{ij}/m_{\tilde{l}}^2$
contributions to LFV are clearly 
correlated.
We may explicitly write the rate of $l_i\to l_j\gamma$ 
in the very suggestive 
form \cite{hepph0702050},
\bea
BR(l_i\to l_j \gamma)\simeq \frac{\alpha^3}{G_F^2}
\frac{m_{\tilde{q}}^4}{M_S^8}|(\delta^d_{ij})_{RR}|^2
\tan^2\beta
\eea 
where $m_{\tilde{q}}$ is the average squark mass and $M_S$ is the 
typical SUSY scale. 

We also need to consider the RGE effects of the CKM mixings in the 
left-handed down squark matrices. These effects can be approximated
as,
\bea
(\delta^d_{ij})_{LL}\approx
-\frac{1}{8\pi^2} Y_t^2\, V^*_{ti}V_{tj}\frac{(3 m_0^2+A_0^2)}{m_{\tilde{q}}^2}
\ln\frac{M^*}{M_W}
\eea
Here $m_0$ is the 
universal scalar mass, $A_0$ the universal A-term and $M^*$
is the scale at which the flavour blind soft SUSY
breaking is communicated.
These effects may be quite important in the 
$\delta^d_{LL}\delta^d_{RR}$ contribution to 
eq.~(\ref{Rg}) due to the typically
large value of $a_4\approx -100\,a_1$. We shall take 
a minimal approach and assume
$M^*$ to be the GUT scale $\sim 2\times 10^{14}$ GeV, in which case 
the mass insertions are of the order, 
$(\delta^d_{23})_{LL}\sim \lambda^2$ and 
$(\delta^d_{13})_{LL}\sim \lambda^3$. Larger values of $M^*$ will
lead to an even more restricted parameter space for 
$(\delta^d_{13,23})_{RR}$. 

If we assume that the second term of eq.~(\ref{Rg}) dominates then 
we may derive the relation,
\bea
\left|\frac{R_s}{R_d}\right|^2
&\approx& 
\left|\frac{a_4^s}{a_4^d}\right|^2
\left|\frac{(\delta^d_{23})_{LL}}{(\delta^d_{13})_{LL}}\right|^2
\left|\frac{(\delta^d_{23})_{RR}}{(\delta^d_{13})_{RR}}\right|^2
\nonumber\\
&\approx& 
\lambda^2
\frac{BR(\tau\to\mu\gamma)}{BR(\tau\to e\gamma)}
\eea
here we have made use of the ratio, 
$|a_4^s|/|a_4^d|\approx V_{td}^2/V_{ts}^2\approx \lambda^2$, as 
derived in appendix A. 
This relation differs from that derived in \cite{hepph0702050}
where it was assumed that, $a_4^s/a_4^d = 1$. In appendix A we show 
the full form of the functions $a_1$ and $a_4$ where it is shown
that the correct ratio is 
$|a_4^s|/|a_4^d|\approx V_{td}^2/V_{ts}^2\approx \lambda^2$. As a result
the allowed parameter space for the mass insertion 
$(\delta^d_{13})_{RR}$ is suppressed relative to $(\delta^d_{23})_{RR}$
by a factor $\lambda^2$ as shown above.

\section{Numerical results and Discussion}

In this section we shall show the results of numerical 
calculations of our generic SUSY analysis of the 
allowed parameter space for the mass insertions $(\delta^d_{13,23})_{RR}$.
We then make use of this allowed parameter space to study 
the correlation of FCNCs and LFV rates. 

For the numerical analysis we fixed the values of $\tan\beta=10$,
$M_S= m_0=M_{1/2}=m_{\tilde{q}}=(300,\,500)$ GeV, $A_0=0$
and the ratio $x\equiv m_{\tilde{g}}^2/m_{\tilde{q}}^2=1$.
Taking these values we scan over
$(\delta^d_{23,13})_{RR}$ and 
require fits to the values of $\rho_q$ and 
$\phi_q^{\rm NP}$ as described in the previous section.

\begin{figure}[h]
 \includegraphics[width=6truecm,clip=true]{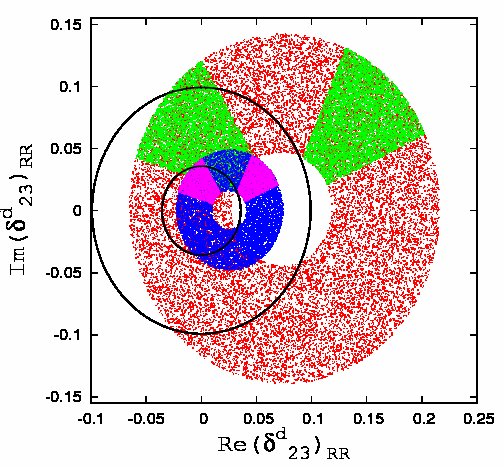}
 \caption{Allowed parameter space for the mass insertion 
$(\delta^d_{23})_{RR}$, with 
$m_{\tilde{q}}=300$(blue/pink) 
and 500(red/green) GeV respectively. Red/Blue points are constrained 
by the measurement of 
$\Delta M_s$ while green/pink points have the extra constraint
from the measurement of $\phi_s$ both at the $95\%$ C.L. 
Solid black lines
show the respective constraints for each value of 
$m_0$ from bounds on $\tau\to\mu\gamma$
from Belle \cite{hepex0609049}.}
 \label{fig1}
 \end{figure}
 
In fig.~\ref{fig1} we show the allowed parameter space for the 
mass insertion $(\delta^d_{23})_{RR}$ as dictated by the constraints
imposed from $\Delta M_s$(red/blue) and $\phi_s$(green/pink). 
The allowed regions form rings in the 
complex $(\delta^d_{23})_{RR}$ plane with wedges cut into it 
corresponding to allowed regions from the measurement of $\phi_s$. 
It is clear therefore that the recent measurements of $\phi_s$ 
substantially restrict the allowed parameter space and provide 
important information. In the SUSY-GUT scenario the same mass insertion
also contributes to the LFV decay $\tau\to\mu\gamma$, the resulting 
bounds from which can be seen by the solid black line
shown in fig~\ref{fig1}. The two black lines correspond to the 
values $m_0=300,\,500$ GeV and clearly exclude considerable regions
of the $(\delta^d_{23})_{RR}$ parameter space. In fact
one of the two allowed regions is strongly disfavoured in the 
SUSY-GUT scenario.

\begin{figure}[h]
 \includegraphics[width=6truecm,clip=true]{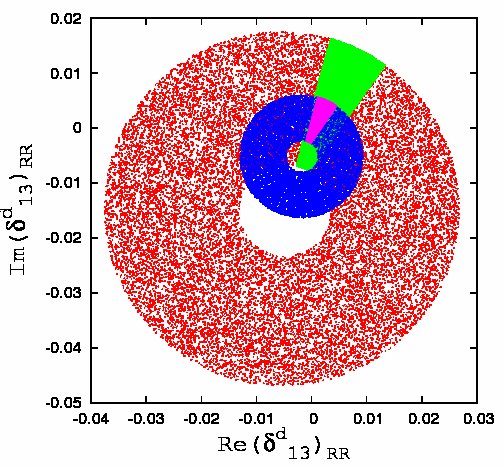}
 \caption{ Allowed parameter space for the mass insertion 
$(\delta^d_{13})_{RR}$, with 
$m_{\tilde{q}}=300$(small) 
and 500(large) GeV respectively. Red/Blue points are constrained 
by the measurement of 
$\Delta M_d$ while green/pink points have the extra constraint
from the measurement of $\phi_d$ both at the $95\%$ C.L..}
 \label{fig2}
 \end{figure}

Fig.~\ref{fig2} shows the respective 
allowed parameter space of the mass insertion 
$(\delta^d_{13})_{RR}$ constrained by the measurements 
of $\Delta M_d$ and $\phi_d$.
For $(\delta^d_{13})_{RR}$ we can see 
that the measurement of $\phi_d$ has reduced the allowed 
parameter space to a small slice in the complex $(\delta^d_{13})_{RR}$
plane. Increasing the 
size of $m_{\tilde{q}}$ leads to an increase in the allowed
parameter space due to a suppression of the functions 
$a_{1,4}\sim 1/m^2_{\tilde{q}}$.

\begin{figure}[h]
 \includegraphics[width=6truecm,clip=true]{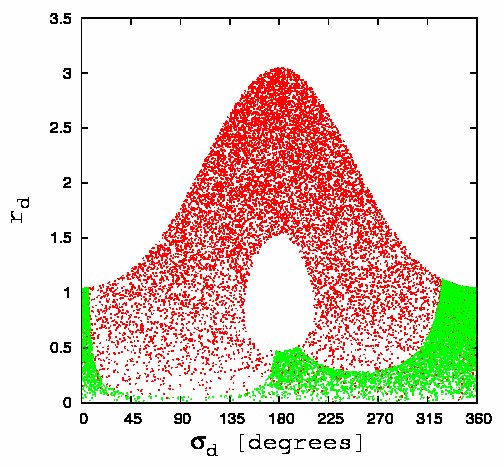}
 \caption{ Allowed region of $r_d-\sigma_d$
parameter space for the NP contribution
to $B_d$ mixing. Red points agree with the $\Delta M_d$ constraint
while green points also agree with that of $\phi_d$
both at the $95\%$ C.L..}
 \label{fig3}
 \end{figure}

\begin{figure}[h]
 \includegraphics[width=6truecm,clip=true]{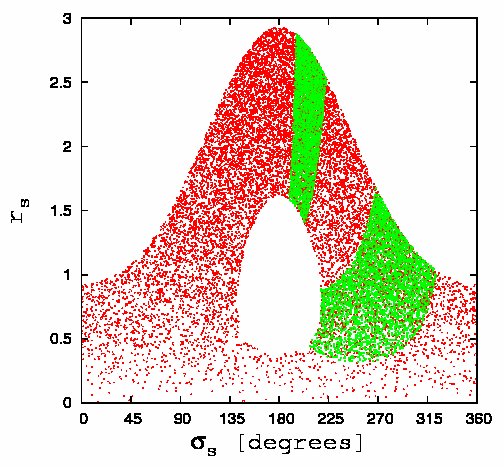}
 \caption{ The allowed region of the $r_s-\sigma_s$ parameter 
space for the NP contribution
to $B_s$ mixing. Points in Red agree with the constraint
from $\Delta M_s$, while green points also agree with the 
$\phi_s$ measurement both at the $95\%$ C.L.. }
 \label{fig4}
 \end{figure}

From these allowed regions of parameter space we can plot
the corresponding allowed regions of the $r_q - \sigma_q$
parameter space. These are shown in fig.~\ref{fig3} 
and fig.~\ref{fig4}. 
Without the restrictions from the CP phases $\phi_{d,s}$
the allowed regions of parameter space form loops in the 
$r_q - \sigma_q$ plane. The constraints from the CP phases
represent slices as shown in \cite{hepph0703214,hepph0604249}.
The accurate measurement of $\phi_d$ leaves the allowed 
region of parameter space as a thin slice shown by the 
green points in fig.~\ref{fig3}.
On the other hand, the present
measurement of $\phi_s$ has a two fold ambiguity shown by the 
two green wedges in fig.~\ref{fig4}.

Next we would like to discuss the implications of this 
allowed parameter space on predictions for the LFV rates
of $\tau\to\mu\gamma$ and $\tau\to e\gamma$. We have shown 
that in a SUSY GUT the mass insertions associated with the 
off-diagonal down squark mass matrix elements and  those
of the sleptons are clearly related. This relation leads to 
a strong correlation between the $B_q$ mixing and LFV rates.
Here we would like to expose this correlation in making 
realistic predictions for LFV rates and to extract information
about the NP CP phase.

\begin{figure}[h]
 \includegraphics[width=6truecm,clip=true]{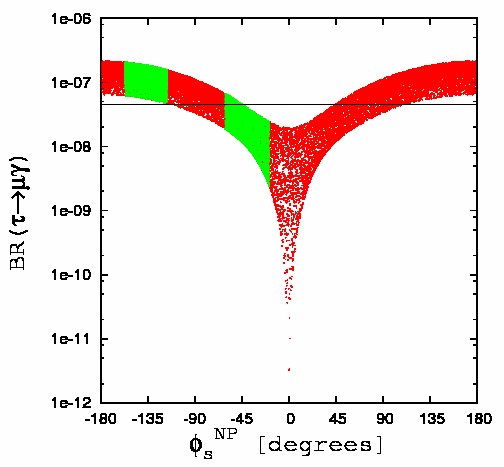}
 \caption{ Predictions for BR$(\tau \to \mu \gamma)$ from 
constraints on $B_s$ mixing. 
The red points conform to the $\Delta M_s$ constraint while
green points show the additional constraint from $\phi_s$,
both at $95\%$ C.L.
The present experimental bound is shown 
by the solid black horizontal line.}
 \label{fig5}
 \end{figure}

Let us first consider the 23 sector. 
Fig.~\ref{fig5}
shows the predicted rate for $\tau\to\mu \gamma$ plotted
against the NP phase $\phi_s^{\rm NP}$.
The possibility of $\phi_s^{\rm NP}=0$ is excluded at the 
3$\sigma$ level \cite{Bona:2008jn}, which also excludes 
the lowest rates for $\tau\to\mu\gamma$ at the 
centre of the plot where $\phi_s^{\rm NP}\to 0$.
The general feature of the plot is that larger LFV rates
are predicted for larger NP phases. 
Without the input of $\phi_s$ this plot is symmetric in 
$\phi_s^{\rm NP}\to -\phi_s^{\rm NP}$ and extends for all 
$\phi_s^{\rm NP}\in (-\pi,\pi)$.
The inclusion of the $\phi_s$ constraint drastically
reduces the allowed parameter space and the resulting prediction
space, as shown in fig.~\ref{fig5}.
These predictions are independent of the choice of 
$M_S=m_{\tilde{q}}$ due to the $1/M^2_{S}$ 
dependence of both BR$(l_i\to l_j\gamma)$
and the functions $a_{1,4}$ shown in appendix A.
There are two allowed regions, one with large NP phase and 
one with smaller NP phase. From fig.~\ref{fig5} it is clear 
that the upper bound on $\tau\to\mu\gamma$ disfavours 
the large phase solution. 
For the second region, there is a lower bound on the 
BR$(\tau\to\mu\gamma)\gtrsim 3\times 10^{-9}$.
This second allowed region is just below 
the present experimental bounds for the branching ratio,
BR$(\tau\to\mu\gamma)< 6.8\times 10^{-8}$ BaBar 
\cite{hepex0502032,hepex0508012} 
and BR$(\tau\to\mu\gamma)< 4.5\times 10^{-8}$ Belle \cite{hepex0609049}. 
From these plots it is clear that improvements of the bound on
$\tau\to\mu\gamma$ in conjunction with an improved measurement of
$\phi_s$ will provide a much stricter test of the 
SUSY GUT scenario. 

\begin{figure}[h]
 \includegraphics[width=6truecm,clip=true]{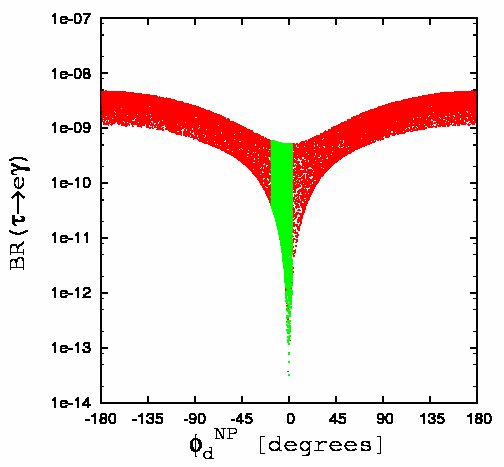}
 \caption{ Predictions for BR$(\tau \to e \gamma)$ from 
constraints on $B_d$ mixing. }
 \label{fig6}
 \end{figure}

In the 13 sector, we can see from fig~\ref{fig2}, 
that the allowed parameter space for the mass insertion
$(\delta^d_{13})_{RR}$ is far more restricted. This is due to the 
smaller mass difference in the $B_d$ system, as a result the functions
$a_1$ and $a_4$ are enhanced by the ratio, 
$\Delta M_d/\Delta M_s\approx V_{td}^2/V_{ts}^2\sim \lambda^2$,
see appendix A. This restriction leads to a much more suppressed
prediction for the rates of $\tau \to e \gamma$. These rates are plotted
in fig.~\ref{fig6} against the allowed values of the 
NP phase $\phi_d^{\rm NP}$. 
Including the constraints from $\phi_d$ the predicted values are in the 
region of $10^{-14}-10^{-9}$. 
The present experimental bounds for the branching ratio
are BR$(\tau\to e\gamma)< 1.1\times 10^{-7}$ BaBar 
\cite{hepex0502032,hepex0508012} 
and BR$(\tau\to e\gamma)< 1.2\times 10^{-7}$ Belle \cite{hepex0609049}. 
Looking at fig.~\ref{fig6} we can see that the allowed region 
of the $(\delta^d_{13})_{RR}$ parameter space predicts rates
well below the present experimental bounds. Therefore the present bounds 
have no impact on the allowed parameter space. 

\begin{figure}[ht]
 \includegraphics[width=6truecm,clip=true]{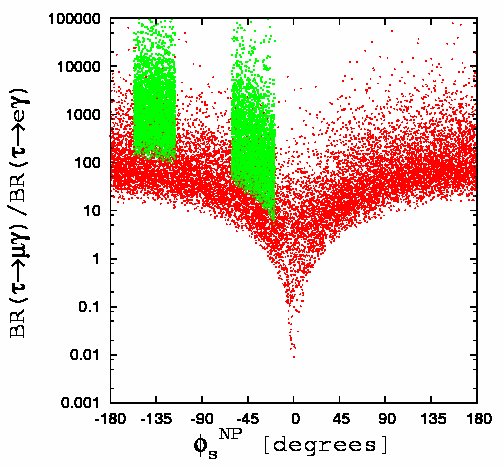}
 \caption{ 
BR$(\tau \to \mu \gamma)$/BR$(\tau \to e \gamma)$ from 
constraints on $B_s$ and $B_d$  mixing. Red points are 
from the $\Delta M_{d,s}$ constraint only, green points
are for both $\Delta M_{d,s}$ and $\phi_{d,s}$ constraints.
}
 \label{fig7}
 \end{figure}

The ratio of the branching fraction of $\tau\to\mu\gamma$ and
$\tau\to e\gamma$ is less dependent on the SUSY parameter
space and so it is interesting to consider the allowed
size of this ratio. Using the parameter space of
$(\delta^d_{13})_{RR}$ and $(\delta^d_{23})_{RR}$ allowed
by the constraints of $\rho_{d,s}$ and $\phi_{d,s}^{\rm NP}$
we can make predictions for
the ratio BR$(\tau \to \mu \gamma)$/BR$(\tau \to e \gamma)$.
Fig.~\ref{fig7} shows the resulting plot of 
BR$(\tau \to \mu \gamma)$/BR$(\tau \to e \gamma)$ plotted
against the prediction for the NP CP phase $\phi_s^{\rm NP}$.
We can see that the numerical results show this
ratio ranging from $~0.01$ up to tens of thousands.
Applying the constraints from the CP phases $\phi_{d,s}$
implies that this ratio must lie in the region $\gtrsim 10$.
This large ratio is again due to the suppression of the 
$(\delta^d_{13})_{RR}$ allowed parameter space due to 
the smallness of the mass difference in the $B_d$ system
as opposed to the larger mass difference in the $B_s$ system.
Here our results disagree with those presented in \cite{hepph0702050}
due to their erroneous assumption of the ratio $a_i^s/a_i^d=1$.
 
Let us now look at a specific example of a simple SUSY GUT.
The example we pick is an SO(10) inspired model as introduced
in \cite{hepph0209303,hepph0205111}. In this model 
there are two Higgs representations one for the up/neutrino
and down/charged lepton Yukawa couplings separately. The neutrino
Yukawa coupling is then related to the up quark Yakawa, where two 
limiting cases are possible; (i) $Y_{\nu}$ has CKM like mixing, 
(ii) $Y_{\nu}$ has MNS like mixing. It is assumed that SO(10)
breaks to SU(5) at the scale $M_{10}$ and further breaks
to the SM gauge group at the scale $M_{\rm GUT}$. Assuming that the 
flavour blind SUSY breaking occurs at the scale 
$M_*=M_{10}=10^{17}$ GeV, then the mass insertions can be written 
as,
\bea
(\delta^d_{ij})_{RR}&\simeq& -\frac{3}{8\pi^2}
y_t^2\,V^*_{j3}V_{i3}
\ln \frac{M_{10}}{M_{\rm GUT}}\\
(\delta^l_{ij})_{LL}&\simeq& -\frac{3}{8\pi^2}
y_t^2\,V^*_{j3}V_{i3}
ln \frac{M_{10}}{M_{R}}
\eea
where the matrix $V$ is $V_{CKM}$ and $U_{MNS}$ for case
(i) and (ii) respectively. The scale $M_R\sim 10^{15}$ GeV 
is the scale of the right-handed neutrinos.
Case (i) produces relatively small 
LFV rates with $(\delta^l_{23})_{LL}\sim 0.008$ and 
$(\delta^d_{23})_{RR}\sim 0.003$. The present bound on
the rate of $\tau\to\mu\gamma$ leads to the weak bound
$m_0\gtrsim 200$ GeV. Alternatively the contribution to $B_s$ mixing
is also very small, for $m_0\gtrsim 200$ GeV we have too small a 
value $|R_s|\lesssim 0.2$.
Case (ii) produces large 
LFV rates with $(\delta^l_{23})_{LL}\sim 0.25$ and 
$(\delta^d_{23})_{RR}\sim 0.08$. Here the present bound on
the rate of $\tau\to\mu\gamma$ leads to the strong bound
$m_0\gtrsim 1000$ GeV. 
Here again the contribution to $B_s$ mixing
is very small, for $m_0\gtrsim 1000$ GeV we once again have too small a
value $|R_s|\lesssim 0.2$.

\section{Summary}

In this work we have re-examined the constraints imposed on the 
parameter space of NP contributions to $B_s$ and $B_d$ mixing
in the light of recent measurements of $\Delta M_s$ and its 
associated CP phase $\phi_s$. 
These constraints were
then imposed to make a generic SUSY
analysis of the allowed parameter space of 
the mass insertions $(\delta^d_{13})_{RR}$ and $(\delta^d_{23})_{RR}$.

In SUSY GUTs there is a deep underlying connection between (s)quarks and 
(s)leptons. 
In such a theory we should therefore
expect that the flavour mixings observed in the quark and lepton 
sectors are correlated. Assuming that such a SUSY GUT exists, we 
have re-examined the correlation between LFVs and FCNCs. 

From our numerical analysis we have found that the 
allowed parameter space for the mass insertion
$(\delta^d_{13})_{RR}$ is particularly restricted by 
the present measurements of $\Delta M_d$ and $\phi_d$.
This is due to the small mass difference in the $B_d$ system
compared to the $B_s$ system.
As a result, the predicted branching ratio for $\tau\to e\gamma$
is particularly suppressed. On the other hand, the present
experimental determination of $\Delta M_s$ and particularly
$\phi_s$ are not yet so restrictive on the allowed values of
$(\delta^d_{23})_{RR}$. The larger mass difference in the
$B_s$ system also means that the predicted values of 
the branching ratio of $\tau\to\mu\gamma$ are much larger
and close to the present bounds from the B factories.
The main conclusion of this work is that  
the recent measurement of $\phi_s$ substantially 
restricts the allowed parameter space and excludes the  
region where the lowest rates for 
$\tau\to\mu\gamma$ are to be found, leading to a lower
bound of $\sim 3\times 10^{-9}$ for $\tan\beta=10$. 
Future experimental improvement of the bound on the decay rate 
$\tau\to\mu\gamma$ and the measurement of $\phi_s$ will lead to a 
strong test of the SUSY-GUT scenario.
We have also considered a specific example of a SUSY SO(10) GUT
and found there to be significant tension between the constraints
from FCNCs and LFV decay rates.


\appendix

\section{$B_{d,s}^0-\bar{B}_{d,s}^0$ mixing in SUSY models}


To examine the contributions from new physics 
we may write the $\Delta B=2$ effective Hamiltonian in the following form,
\be
{\mathcal{H}}_{eff}^{\Delta B=2}=\sum_{i=1}^{5}C_i(\mu)Q_i(\mu)+ h.c.
\ee
The operators are defined as,
\bea
Q_1&=&\bar{q}_L^{\alpha}\gamma_{\mu}b_L^{\alpha}\,
\bar{q}_L^{\beta}\gamma^{\mu}b_L^{\beta}\\
Q_2&=&\bar{q}_R^{\alpha}b_L^{\alpha}\,
\bar{q}_R^{\beta}b_L^{\beta}\\
Q_3&=&\bar{q}_R^{\alpha}b_L^{\beta}\,
\bar{q}_R^{\beta}b_L^{\alpha}\\
Q_4&=&\bar{q}_R^{\alpha}b_L^{\alpha}\,
\bar{q}_L^{\beta}b_R^{\beta}\\
Q_5&=&\bar{q}_R^{\alpha}b_L^{\beta}\,
\bar{q}_L^{\beta}b_R^{\alpha}
\eea
We shall also need to define the hadronic matrix elements of the 
above operators such that,
\bea
\langle B_q | Q_1 |\bar{B}_q \rangle &=& -\frac{1}{3}M_{B_q}f_{B_q}^2 B_1(\mu)\\
\langle B_q | Q_2 |\bar{B}_q \rangle &=& 
\frac{5}{24}R_{B_q} M_{B_q}f_{B_q}^2 B_2(\mu)\\
\langle B_q | Q_3 |\bar{B}_q \rangle &=& 
-\frac{1}{24}R_{B_q} M_{B_q}f_{B_q}^2 B_3(\mu)\\
\langle B_q | Q_4 |\bar{B}_q \rangle &=& 
-\frac{1}{4}R_{B_q} M_{B_q}f_{B_q}^2 B_4(\mu)\\
\langle B_q | Q_5 |\bar{B}_q \rangle &=& 
-\frac{1}{12}R_{B_q}M_{B_q}f_{B_q}^2 B_5(\mu)
\eea
where we have defined 
$R_{B_q}=\left(\frac{M_{B_q}}{m_b(\mu)+m_q(\mu)}\right)^2$.
The values of the bag parameters have been calculated on the lattice 
and can be found in \cite{hepph0110091}.

%
The dominant SUSY contribution is from the gluino, which gives
the following Wilson coefficients 
\cite{hepph0311361,Gabbiani:1988rb,
Hagelin:1992tc,Gabrielli:1995bd,Gabbiani:1996hi,hepph0112303}
\bea
&&\hspace{-0.7cm}
C_1^{\tilde{g}}(M_S)=-\frac{\alpha_s^2(M_S)}{216\,m_{\tilde{q}}^2}
\nonumber\\
&\hspace{-0.2cm}\times&\hspace{-0.2cm}\left[24 x f_6(x)
+66\tilde{f}_6(x)\right](\delta_{q3}^d)^2_{LL}  \\
&&\hspace{-0.7cm}
C_4^{\tilde{g}}(M_S)=-\frac{\alpha_s^2(M_S)}{216\,m_{\tilde{q}}^2}
\nonumber\\
&\hspace{-0.2cm}\times&\hspace{-0.2cm}
\left[504 x f_6(x)
-72\tilde{f}_6(x)\right](\delta_{q3}^d)_{LL}(\delta_{q3}^d)_{RR} \\
\hspace{-0.2cm}&&\hspace{-0.7cm}
C_5^{\tilde{g}}(M_S)=-\frac{\alpha_s^2(M_S)}{216\,m_{\tilde{q}}^2}
\nonumber\\
&\hspace{-0.2cm}\times&\hspace{-0.2cm}
\left[24 x f_6(x)
+120\tilde{f}_6(x)\right](\delta_{q3}^d)_{LL}(\delta_{q3}^d)_{RR} 
\eea
Here only the LL and RR terms have been kept as the contributions
from RL and LR are suppressed. The above Wilson coefficients are 
evaluated at the scale 
$M_S=(M_{\tilde{g}}+M_{\tilde{q}})/2=(\sqrt{x}+1)M_{\tilde{q}}/2$.

These Wilson coefficients need to be renormalization group (RG) 
evolved to the scale of the
bottom quark. Using the magic numbers from \cite{hepph0112303}, we have,
\be
C_{\alpha}(\mu_b)
=\sum_i \sum_{\beta}\, 
(b_i^{(\alpha,\beta)}+\eta\, c_i^{(\alpha,\beta)})
\,\eta^{a_i}\,C_{\beta}(M_S)
\ee
with $\eta=\alpha_s(M_S)/\alpha_s(\mu_t)$. Let us now write the 
$B_s-\bar{B}_s$ mixing parameter 
\bea
\Delta M_q\equiv 2|M_{12}^q|&\hspace{-0.1cm}=&\hspace{-0.1cm}
2|M_{12}^{q,{\rm SM}}(1+R_q)|\nonumber\\
&\hspace{-0.1cm}=&\hspace{-0.1cm}
\Delta M_{q}^{\rm SM}|1+R_q^{\tilde{g}}+\ldots|
\eea
where $R_q\equiv r_q e^{i \sigma_q}=M_{12}^{q,{\rm NP}}/M_{12}^{q,{\rm SM}}$ 
parameterizes the 
contribution from new physics. We consider the dominant effect
to come from the gluino 
$R_q^{\tilde{g}}=M_{12}^{q,\tilde{g}}/M_{12}^{q,{\rm SM}}$.
From the above effective Hamiltonian, Wilson coefficients, operator
matrix elements and RG evolution we can write,
\bea
R^{\tilde{g}}_q&\hspace{-0.1cm}=&\hspace{-0.1cm}
a^q_1(m_{\tilde{g}},x)\,
\left[(\delta_{q3}^d)^2_{RR}+(\delta_{q3}^d)^2_{LL}\right]
\nonumber\\
&&\hspace{-0.2cm}+a^q_4(m_{\tilde{g}},x)\,
(\delta_{q3}^d)_{LL}(\delta_{q3}^d)_{RR}+\ldots
\eea
have we have only kept the LL, RR terms as they will dominate.
The complete form of the functions $a_1(m_{\tilde{g}},x)$ and
$a_4(m_{\tilde{g}},x)$ are as follows,
\bea
M_{12}^{q,{\rm SM}}\,a^q_1(m_{\tilde{g}},x)&\hspace{-0.1cm}=&\hspace{-0.1cm}
\frac{-\alpha_s^2(M_S)}{216\,m_{\tilde{q}}^2}
\mathcal{A}\nonumber\\
&\hspace{-5cm}\times&\hspace{-2.5cm}
\frac{1}{3}m_{B_s}f_{B_s}^2 B_1(\mu)
\sum_i (b_i^{(1,1)}+\eta\, c_i^{(1,1)})\,\eta^{a_i}\label{a1}\\
M_{12}^{q,{\rm SM}}\,a^q_4(m_{\tilde{g}},x)&\hspace{-0.1cm}=&\hspace{-0.1cm}
\frac{-\alpha_s^2(M_S)}{216\,m_{\tilde{q}}^2}
m_{B_s}f_{B_s}^2
R_{B_q}\nonumber\\
&&\hspace{-3cm}
\times\left[\,\mathcal{B}
\sum_i (b_i^{(4,4)}+\eta\, c_i^{(4,4)})\,\eta^{a_i}\right.
\frac{1}{4} B_4(\mu)\nonumber\\
&&\hspace{-2.5cm}\left.
+\mathcal{A}
\sum_i (b_i^{(4,5)}+\eta\, c_i^{(4,5)})\,\eta^{a_i}\right.
\frac{1}{4} B_4(\mu)\nonumber\\
&&\hspace{-2.5cm}\left.+\mathcal{B}
\sum_i (b_i^{(5,4)}+\eta\, c_i^{(5,4)})\,\eta^{a_i}
\frac{1}{12} B_5(\mu)
\right.\nonumber\\
&&\hspace{-2.5cm}\left.
+\mathcal{A}
\sum_i (b_i^{(5,5)}+\eta\, c_i^{(5,5)})\,\eta^{a_i}
\frac{1}{12} B_5(\mu)
\right]\label{a4}
\eea
Where we have defined, $\mathcal{A}=24\, x\, f_6(x)+66\,\tilde{f}_6(x)$ and 
$\mathcal{B}=504 \, x \, f_6(x)-72\,\tilde{f}_6(x)$ where,
\bea
f_6(x)
&\hspace{-0.2cm}=&\hspace{-0.2cm}
\frac{6(1+3x)\ln x+x^3-9x^2-9x+17}{6(x-1)^5}  \\
\tilde{f}_6(x)
&\hspace{-0.2cm}=&\hspace{-0.2cm}
\frac{6x(1+x)\ln x-x^3-9x^2+9x+1}{3(x-1)^5}
\eea
and $x=m_{\tilde{g}}^2/m_{\tilde{q}}^2$.

The dominant contributions to the above functions come 
from $b_1^{(1,1)}=0.865$,  $b_4^{(4,4)}=2.87$,
 $b_4^{(4,5)}=0.961$,  $b_4^{(5,4)}=0.09$, and
 $b_5^{(5,5)}=0.863$, with $a_{i}=(0.286,-0.692,0.787,-1.143,0.143)$.

So we can write the approximate functions as,
\bea
M_{12}^{q,{\rm SM}}\,a^q_1(m_{\tilde{g}},x)&\approx&
\frac{-\alpha_s^2(M_S)}{216\,m_{\tilde{q}}^2}
\frac{1}{3}m_{B_s}f_{B_s}^2 B_1(\mu)\nonumber\\
&&\hspace{-2cm}\times\,\mathcal{A}
\,0.865\,\eta^{0.286}\label{a1sim}\\
M_{12}^{q,{\rm SM}}\,a^q_4(m_{\tilde{g}},x)&\approx&
\frac{-\alpha_s^2(M_S)}{216\,m_{\tilde{q}}^2}
R_{B_q}
m_{B_s}f_{B_s}^2 
\nonumber\\
&&\hspace{-3.1cm}\times
\left[\left[\mathcal{B}
\,2.87\,\eta^{-1.143}
+\mathcal{A}
\,0.961\,\eta^{-1.143}\right]\frac{1}{4}B_4(\mu)\right.\nonumber\\
&&\left.\hspace{-3.1cm}+\,\left[\mathcal{B}
\, 0.09\,\eta^{-1.143}
+\mathcal{A}
\,0.863\,\eta^{0.143}\right]\frac{1}{12}B_5(\mu)\right]
\label{a4sim}
\eea


We can see that the values of the functions $a_{1,4}$
are not the same for the $B_s$ and $B_d$ systems.
From eq.~(\ref{a1},\ref{a4}) 
we can approximate the ratio of these functions as,
\be
\frac{|a_1^s|}{|a_1^d|}\approx \frac{|a_4^s|}{|a_4^d|}\sim
\frac{\Delta M_d^{\rm SM}}{\Delta M_s^{\rm SM}}
\sim \frac{V_{td}^2}{V_{ts}^2}\sim \lambda^2
\ee
so that the value of the functions $a_1$ and $a_4$ are enhanced
in the $B_d$ relative to those derived in the $B_s$ system.
This enhancement will lead to a more constrained parameter
space of the 13 mass insertion relative to the 23 sector.
It is also important to notice that $a_1$ and $a_4$ are complex 
parameters. This is particularly important in the case of the $B_d$
system where the phase of $M_{12}^{d,\rm SM}$, 
$\phi_d^{\rm SM}=2\beta$, is large.

 \null\noindent
 {\bf Acknowledgment:}~ JKP would like to thank E.~Kou for 
useful discussions concerning the gluino contribution to $B_s$
mixing. HHZ thanks Prof~Y.~P.~Kuang and Prof~Q.~Wang 
for their encouragement. This work is supported by the 
National Natural Science Foundation of China under 
Grant No. 10747165, and Sun Yet-Sen University Science Foundation


\end{document}